\documentclass[sigconf]{acmart}

\renewcommand\footnotetextcopyrightpermission[1]{} % removes footnote with conference information in first column
\usepackage{booktabs} % For formal tables

\usepackage{amssymb}
\usepackage[ruled]{algorithm2e} % For algorithms

\newcommand{\slim}{{\scshape Slim}}
\newcommand{\wmf}{{\scshape wmf}}
\newcommand{\cdae}{{\scshape cdae}}

\newcommand{\mvae}{${\textsc{Mult-vae}}^{\textsc{ pr}}$}
\newcommand{\mdae}{{\scshape Mult-dae}}

\newcommand{\RR}{\mathbb{R}}
\newcommand{\EE}{\mathbb{E}}
\newcommand{\I}{\mathcal{ I}}
\newcommand{\UU}{\mathcal{U}}
\newcommand{\LL}{\mathcal{L}}
\newcommand{\OO}{\mathcal{O}}
\newcommand{\diag}{{\rm diag}}
\newcommand{\dMat}{{\rm diagMat}}

\newcommand{\PP}{\hat{P}}
\newcommand{\Blr}{\hat{B}^{\rm (rr)}  }

\newcommand{\Bzero}{\hat{B}^{\rm (0d)}  }
\newcommand{\BzeroXXno}{\hat{B}^{\rm (0d)}  }
\newcommand{\BzeroXX}{\hat{B}^{\rm (0d)}_{(X=Y)}  }
\newcommand{\Bsparse}{\hat{B}^{\rm (sparse)}  }
\newcommand{\Wu}{w^{\rm(U)}  }
\newcommand{\Wi}{w^{\rm(I)}  }
\newcommand{\Bs}{\hat{B}^{\rm (scaled)}  }

\newcommand{\Bw}{\hat{B}^{\rm (weighted)}  }
\newcommand{\pop}{{\rm pop}}
\newcommand{\cor}{{\rm cor}}

\begin{document}
\title{Collaborative Filtering via High-Dimensional Regression}

\author{Harald Steck}
\affiliation{%
  \institution{Netflix}
  \city{Los Gatos}
  \state{California}
}
\email{hsteck@netflix.com}

\begin{abstract}
While the \slim{} approach \cite{ning11} obtained high ranking-accuracy in many experiments in the literature, it is also known for its high computational cost of learning its parameters from data. For this reason, we focus  in this paper on  variants of high-dimensional  regression problems that have \emph{closed-form} solutions. Moreover, we motivate a re-scaling rather than a re-weighting approach for dealing with  biases regarding item-popularities in the data. We also discuss properties of the sparse solution, and outline a computationally efficient approximation.
 In  experiments on three publicly available data sets, we observed not only extremely reduced training times, but also significantly improved ranking accuracy compared to \slim{}. Surprisingly, various state-of-the-art models, including deep non-linear autoencoders, were also outperformed on two of the three data sets in our experiments, in particular  for  recommendations with highly personalized relevance.
\end{abstract}

\maketitle

\section{Introduction}
\slim{} \cite{ning11} is a linear-regression model with several constraints, and achieved competitive recommendation-accuracy in the literature, despite its simplicity. These constraints, however, render the training computationally expensive.
In this paper, we simplify this approach and discuss several extensions. This article summarizes several extensions of our short paper \cite{steck19a}. The main contributions are as follows:
\begin{itemize}

\item Compared to \slim, we dropped  the  L1-norm regularization-term and the non-negativity constraint  on the learned weights for computational efficiency. Surprisingly, we observed considerable improvements in ranking-accuracy, even outperforming other competing models, like deep non-linear autoencoders on two of the three data sets in our experiments. 

\item  In \slim{} \cite{ning11} and its variants \cite{levy13, sedhain16},  the learning problem was decomposed into independent tasks, one for each item, which is computationally very costly, even though it is embarrassingly parallel. We show that it is possible to solve a modified problem with a \emph{single} closed-form solution  (Section \ref{sec_0d}). Closed-form solutions are the main focus of this paper.

\item Whereas  the constraint of  a zero-diagonal in the learned weight-matrix was dropped in  variants  \cite{levy13, sedhain16} of \slim{} for computational reasons, we found this constraint to be  crucial for improved ranking-accuracy  in our experiments (Sec. \ref{sec_0d}).

\item We show that biases in the data,  in particular adjustments in item-popularities, can be tackled effectively by re-scaling the target-values, rather than re-weighting the errors,  in the presented approach (Section \ref{sec_covshift}).

\item  In Section \ref{sec_sparse}, we discuss that the main advantage of sparse modeling in collaborative filtering may be in the reduction of the computational cost rather than in  improvements in  ranking-accuracy. Sparse modeling may improve recommendation accuracy, however, by reducing the occurrence of trust-busters in the sense of generally popular items that are unrelated to the user's past user-item interactions. While training sparse models can be computationally expensive, we outline an efficient approximate approach.

\item The learned weight-matrix of the regression-model may also be interpreted as the  item-item similarity-matrix in a neighborhood-based approach (e.g., see \cite{sedhain16}). The closed-form solution in this paper reveals that  the conceptually  correct similarity matrix is determined by the \emph{inverse} of the given item-item (or user-user) data-matrix (Section \ref{sec_0d}), which is in stark contrast to existing approaches (e.g., see \cite{koen14, volkovs15} and references therein).

\end{itemize}
The  computational cost is discussed in Section \ref{sec_complexity}, where we also highlight the greatly reduced  training-time observed in our experiments. 
We finish this paper with a summary of the experimental set-up (Section \ref{sec_exp}). 
Related work is discussed in each of the sections regarding the different variants outlined in this paper.
\section{Approach: Preliminaries}
\label{sec_prelim}
Let the training data be given in the form of two user-item interaction matrices $Y, X \in \RR^{|\UU|\times |\I|}$, where $\UU$ and $\I$ are the sets of users and items in the training data, respectively, and $|\cdot|$ denotes the cardinality of a set. These matrices  are typically sparse (unobserved interactions are represented by zero). Observed interactions may be represented by ones (e.g., user has listened to a song) or by continuous values (e.g., the time a user listened to a song). 

We use two matrices $Y, X$ for training, as to allow for the fact that they possibly hold different data: for instance, in practical applications,  $X$ may represent the past user-item interactions, while $Y$ may reflect the future ones, relative to a chosen reference-time. Also, $X$ and $Y$ may hold different kinds of user-item interactions, like product-views and product-purchases on a shopping site.
We also allow for $X=Y$ (as is common in the literature), except for the  approach in  Section \ref{sec_xy0d}.

The  linear model considered in this paper is defined by its  item-item weight-matrix $B\in\RR^{|\I|\times |\I|}$. The  model-predictions are 
\begin{equation}
\hat{Y}= X \hat{B},
\end{equation}
where $\hat{B}$ denotes the weight-matrix estimated from the given data, and $\hat{Y}$ are the predicted scores.\footnote{While $\hat{B}$ captures \emph{pairwise} (item-item) parameters, also \emph{unary} item-intercept-parameters may be included in the model by  appending a column of ones to $X$ and a row (of item-intercepts) to the bottom of $\hat{B}$.  As we did not find significant improvements in our experiments, only results without the item-intercepts are reported. An alternative to learning the item-intercepts as part of $B$, is to simply center each column $i\in \I$ in the training-matrix $Y$ by subtracting its mean $\mu_i$, then learn $\hat{B}$ (without item-intercepts), and add the means back  as to obtain the final scores $\hat{Y}+ \vec{1}\cdot \mu^\top = X\hat{B}+\vec{1}\cdot \mu^\top$, with a column-vector of ones and the vector of means $\mu=(\mu_1,...,\mu_{|\I|})^\top$. The items are finally ranked by these scores.}
The training objective is 
\begin{equation}
\Blr = \arg \min_B ||Y-X\cdot B||_F^2 + \lambda \cdot ||B||_F^2   ,
\label{eq_opt_blr}
\end{equation}
where $||\cdot ||_F$ denotes the Frobenius norm.  L$_{2}$-norm regularization  with hyper-parameter $\lambda>0$ is used to prevent overfitting (ridge regression). We chose this simple training objective, as it allows for \emph{closed-form solutions}, as will be discussed in the remainder of this paper.  The well-known solution of Eq. \ref{eq_opt_blr} is given by 
\begin{eqnarray}
&\Blr &= \PP \cdot X^\top Y     ,   \label{eq_blr} \\
{\rm where}\,\,\,\,\,\,&\PP &= \left ( X^\top X +\lambda \cdot I\right )^{-1}  ,
\label{eq_pp}
\end{eqnarray}
where $I$ denotes the identity matrix. Obviously, if $Y=X$ and  $\lambda=0$, then $\Blr=I$ is the trivial solution, and is hence not useful. This motivates the constraint of a zero diagonal, of which two different variants are discussed in the following section.

\section{Zero Diagonals}
\label{sec_0d}
We first outline the constraint of a zero diagonal in the weight matrix $B$ (as fist introduced in \slim{} \cite{ning11}), and derive the closed-form solution. As a simple alternative (see Section \ref{sec_xy0d}), yet with slightly less accurate predictions, one may split the training data into disjoint matrices $X,Y$. 
\subsection{Zero Diagonal in the Weight Matrix}
\label{secb0d}
As to exclude the trivial solution $B=I$, we now add the constraint that the diagonal of the weight matrix has to vanish (this constraint  was first introduced in  \slim{}  \cite{ning11}, but was dropped in later variants \cite{levy13, sedhain16} for computational efficiency):
\begin{eqnarray}
\Bzero &&= \arg \min_B ||Y-X\cdot B||_F^2 + \lambda \cdot ||B||_F^2 \nonumber\\
&&{\rm s.t.}\,\,\,\,\, \diag(B)=0
\label{eq_opt_bzero}
\end{eqnarray}
\slim{} \cite{ning11} and its variants \cite{levy13, sedhain16}  took advantage of the fact that this  least-squares problem  decomposes into separate least-squares problems, one for each column/item $i \in \I$,
\begin{equation}
||Y-X\cdot B||_F^2 = \sum_{i\in \I} ||Y_{\cdot,i}-X\cdot B_{\cdot,i}||_F^2,
\label{eq_columns}
\end{equation} 
which can be solved independently of each other. Despite embarrassingly parallel computations, this is costly or even prohibitive in domains with a large number of items, e.g., see row (a) in Table \ref{tab_msd}, where the original implementation by the authors \cite{ning11} was used.

This motivated us to derive the closed-form solution of Eq. \ref{eq_opt_bzero},  using  the method of Lagrangian multipliers, which  applies to equality constraints.
We define the vector of Lagrangian multipliers $\gamma=(\gamma_1,...,\gamma_{|\I|})^\top$ and form the Lagrangian:
\begin{equation}
L=||Y-X B||_F^2 + \lambda\cdot||B||_F^2 + 2\cdot\gamma^\top\cdot \diag(B),
\nonumber
\end{equation}
The constrained optimization problem in  Eq. \ref{eq_opt_bzero}  is solved by minimizing this Lagrangian. We hence set its derivative  to zero, which yields the estimate $\Bzero$ after re-arranging terms:
\begin{eqnarray}
\Bzero &=& \PP \cdot \left(  X^\top Y - \dMat(\gamma) \right )\label{eq_bzero_0}\\
&=& \Blr - \PP \cdot \dMat(\gamma) ,
\label{eq_bzero_1}
\end{eqnarray}
where  $\Blr$ and $\PP$  are  given by Eqs.  \ref{eq_blr} and \ref{eq_pp};  $\dMat(\gamma)$ denotes the diagonal matrix with the Lagrangian multipliers $\gamma=(\gamma_1,...,\gamma_{|\I|})^\top$. 
Their values are determined by the constraint $\diag(\Bzero)=0$, which yields
\begin{equation*}
0=\diag(\Bzero) = \diag(\Blr) - \diag(\PP) \odot \gamma,
\end{equation*}
where $\odot$ denotes the elementwise product. It follows that
\begin{equation}
\gamma = \diag(\Blr) \oslash \diag(\PP), \nonumber
\end{equation}
where $\oslash$  denotes the elementwise division of the two vectors on the diagonals of the matrices $\Blr$ and $\PP$ (which is well-defined given that $\PP$ is invertible).
 Substituting this back into Eq. \ref{eq_bzero_1}, yields 
\begin{equation}
\Bzero = \Blr - \PP \cdot \dMat\left(\diag(\Blr) \oslash \diag(\PP)\right) ,
\label{eq_bzero}
\end{equation}
which is the  closed-form solution of Eq. \ref{eq_opt_bzero}.

If $X=Y$, we have $X^\top Y = X^\top X = \PP^{-1}-\lambda\cdot I$ (see Eq. \ref{eq_pp}). This identity, together with Eq. \ref{eq_blr}. is substituted into Eq. \ref{eq_bzero}, which further simplifies  for $X=Y$:
\begin{equation}
\BzeroXX = I - \PP \cdot \dMat \left ( \vec{1} \oslash \diag(\PP) \right),
\label{eq_bzeroxx}
\end{equation}
i.e.,  the inverted matrix $\PP$ (see Eq. \ref{eq_pp}) fully determines  $\BzeroXX$; in fact, the off-diagonal elements of $\BzeroXX$ are obtained by dividing each column $i$ of $\PP$ by its diagonal element $\PP_{ii}$. Hence, even  if $X=Y$, we have that $\BzeroXX$ is an asymmetric matrix in general,  even though   $\PP$ is symmetric. 

Given that $\Bzero$ may also be interpreted as the similarity-matrix in a  neighborhood-based approach, Eqs. \ref{eq_bzeroxx} and \ref{eq_pp} show that the conceptually correct similarity-matrix is asymmetric and is based on the \emph{inverse} of the data Gram-matrix  $X^\top X$. These are two key differences to the similarity-matrices commonly used in neighborhood-based approaches  (e.g., see \cite{koen14, volkovs15} and references therein), which typically employ re-scaled versions of  $X^\top X$ (e.g.,  cosine similarity).  

{\bf Experiments:} Surprisingly, in our experiments we found that $\BzeroXXno$, despite its simplicity, was not only competitive with the various baselines  in rows (a)-(e) in Table \ref{tab_ml20m}, but also outperformed them  in Tables \ref{tab_nflx} and \ref{tab_msd}--in the latter, remarkably by about 20\%.

\subsection{Zero Diagonal in the Data Gram-Matrix}
\label{sec_xy0d}
In this section, we show that the unconstrained regression-problem, see Eqs. \ref{eq_opt_blr}-\ref{eq_pp}, can be a useful approach if the data matrices are (forced to be) \emph{disjoint} in the sense  that each observed user-item-interaction is reflected by either  $Y$ or  $X$ (but not both), i.e.,  $Y \odot X =0$ where $\odot$ denotes the elementwise product (unobserved interactions are represented by 0). If $Y \odot X =0$, then it is easy to see that $\diag(X^\top Y)=0$, i.e., we now have a zero diagonal in the data Gram-matrix $X^\top Y$. While different from the constraint $\diag(B)=0$ in Eq. \ref{eq_opt_bzero}, we show in the following that both are almost the same.

Given that only a single training matrix $Z$ was available in our experiments on publicly available data sets, we created disjoint matrices $Y,X$ by random splits of $Z$, see Appendix for details: as a result, we used  $X^\top X := Z^\top Z$ and $X^\top Y := Z^\top Z - \dMat(\diag(Z^\top Z))$ in our experiments. Substituted into Eqs. \ref{eq_blr} and \ref{eq_pp}, we obtain for $\Blr$:
\begin{equation}
\Blr_{(Z)} = I - \PP\cdot \dMat \left ( \diag(Z^\top Z) +\lambda \right ).
\label{eq_blrzz}
\end{equation}
Note its similarity to Eq. \ref{eq_bzeroxx}, where one can re-write  $\dMat(\vec{1}\oslash \diag(\PP)) =  \dMat \left (\vec{1}\oslash\diag(( Z^\top Z+\lambda I)^{-1}) \right )$ for $Z=X$.

{\bf Experiments:}  While less accurate than $\Bzero$ in our experiments,  even this simple approach (row 2) was competitive with  several of the baselines in rows (a)-(e) in  Tables \ref{tab_ml20m}, \ref{tab_nflx}, and \ref{tab_msd}.
\section{Biased Training-Data}
\label{sec_covshift}
A key challenge  in real-world applications of recommender systems is the removal of the various biases that are present in the data. Several different  approaches have been developed for estimating and removing biases (e.g., \cite{liang16a,liang16b,schnabel16a}), often based on inverse propensity scoring. 
A very prominent bias is due to the fact that the data are  missing not at random \cite{marlin09}. A simple, yet effective, approach is to sample negative user-item interactions when learning the model from training-data that are mainly comprised of positive user-item interactions, as is done in weighted matrix factorization \cite{hu08,pan08,steck10}.  Another useful application is the removal of the popularity-bias in the data, so that the model can learn item-\emph{similarities} that are not tainted by  item-popularities \cite{steck11,mikolov13}
--in the domain of natural language processing this was done by word2vec \cite{mikolov13}.  

In this section, we first outline the \emph{weighted} version of matrix-based regression. As it cannot be solved in closed form in general, we then motivate a \emph{re-scaled} version, which can be solved easily.
\subsection{Weighted Errors}
Allowing for a possibly different weight $W_{u,i}$ regarding the squared error of each user $u$ and item $i$ is the most general weighting scheme: 
\begin{eqnarray}
\Bw = &&\arg \min_B || \sqrt{W} \odot ( Y -X\cdot B)||_F^2 + \lambda \cdot  ||B ||_F^2 \nonumber\\
&&{\rm s.t.}\,\,\,\,\, \diag(B)=0,
\label{eq_weighted}
\end{eqnarray}
where $\odot$ denotes the elementwise product of 
the elementwise square-root of the weighting-matrix $W\in \RR ^{|\UU| \times |\I|}$ concerning the error-matrix $Y -XB$. This problem may be solved by decomposing it into separate columns (like in Eq. \ref{eq_columns}), and solving a weighted linear regression for each column $i \in \I$, analogous to \slim{} \cite{ning11}. A closed-form solution to the linear problem where $B$ is a matrix rather than a vector,  unfortunately does not exist for a  general weighting-matrix $W$. Two important special cases with a closed-form solution are as follows.
\subsubsection{Weighting of Users}
If $W  = \Wu\cdot \vec{1}^\top$ is the outer product of  the vector of user-weights $\Wu\in \RR^{|\UU|}$ and a vector of ones, then
$|| \sqrt{W} \odot ( Y -X\cdot B)||_F^2  = || \dMat(\sqrt{\Wu}) \cdot ( Y -X\cdot B)||_F^2$, and it is easy to see that the solution for $\Bw$ is given by Eq. \ref{eq_bzero} (see also Eqs. \ref{eq_blr} and \ref{eq_pp}) after replacing  $X^\top X$ by $X^\top \dMat(\Wu) X $, and  $X^\top Y$ by $X^\top \dMat(\Wu) Y $. Note that this re-weighting may be done in the data-preprocessing step when these two item-item matrices are generated, prior to the training.
\subsubsection{Weighting of Items}
If $W =  \vec{1} \cdot\Wi{}^\top$ is the outer product of a vector of ones and the vector of item-weights  $\Wi\in \RR ^{|\I|}$,  then 
 $|| \sqrt{W} \odot ( Y -X\cdot B)||_F^2 = || ( Y -X\cdot B)\cdot \dMat(\sqrt{\Wi})||_F^2  = \sum_{i \in \I} \Wi_i \cdot ||Y_{\cdot,i} -X\cdot B_{\cdot,i}  ||_2^2$. The last identity shows that the squared error decomposes into a weighted sum of independent squared errors, one regarding each column $i\in \I$. 
Hence, the optimal solution $\Bw_{\cdot,i}$ for column $i$ is unaffected by  item-weight $\Wi_i$ (assuming that the L2-norm regularization is re-scaled accordingly). Consequently, this simple weighting scheme has \emph{no} effect on the learned model-weights $\Bw$.

For the item-weighting to have an effect, for each item $i$, \emph{different} weights have to be used across users. For instance, this is done in weighted matrix factorization \cite{hu08,pan08,steck10}, where the weight of a user-item interaction depends on the fact whether it was observed or  was missing in the data-matrix. Unfortunately, such a weighting scheme with $W_{u,i}$ does not have a closed-form solution for matrix $\Bw$ in general. 
This motivated us to consider a different approach, which allows for a closed-form solution using only item-weights $\Wi$, as outlined in the following section.
\subsection{Re-scaled Target-Values}
\label{sec_rescaled}
In this section we show the effectiveness of  \emph{re-scaling} the target values $Y$ with item-weights $\Wi\in \RR^{|\I|}$. Note that this avoids the use of weights $W_{u,i}$ that depend on both users and items, which would prevent a closed-form solution. Re-scaling the target values,

\begin{eqnarray}
\Bs && = \arg \min_B ||Y\cdot \dMat(\Wi) -X\cdot B||_F^2 +\lambda \cdot  ||B||_F^2 \nonumber \\
&&{\rm s.t.}\,\,\,\,\, \diag(B)=0,
\label{eq_opt_rescaled}
\end{eqnarray}
may be motivated as follows: let us consider the special case that $Y$ is a binary matrix, reflecting the observed (value 1) and missing (value 0) user-item interactions. For each item $i\in I$, if we re-weight the squared errors (like in Eq. \ref{eq_weighted}) depending on the fact  whether  the user-item-interaction is observed (weight $w_1$) or missing (weight $w_0$) in $Y$, then the leading-order effect is that the trained model will predict a different mean (or intercept) for the item (see also \cite{king01} for logistic regression). For instance, the intercept  will increase as we increase $w_1$ relative to $w_0$. Now, the same leading-order effect can also be achieved by re-scaling (see Eq. \ref{eq_opt_rescaled}) while using only the vector of item-weights $\Wi$ (i.e., using only \emph{one} weight per item): the reason is that, for each item $i\in\I$, $\Wi_i$ re-scales only the observed user-item interactions due to their value of 1 in $Y$, while the missing ones are unaffected by $\Wi_i$ due to their value of 0; hence the mean (or intercept) for each item $i$ can be controlled by re-scaling with weight $\Wi_i$. The difference between re-weighting and re-scaling is that the  errors are quantified in different ways. This may be of concern if the ultimate objective is to (exactly) optimize the weighted squared error. In our case, however, neither one of the training objectives (re-weighted in Eq. \ref{eq_weighted} or re-scaled in Eq. \ref{eq_opt_rescaled}) matches exactly our final goal of optimizing a ranking-metric (on the test data). We hence use the re-scaled objective in Eq. \ref{eq_opt_rescaled}  as a surrogate objective, as it has a closed-form solution. It can be derived easily:
first, we
 substitute $\tilde{Y} := Y\cdot \dMat(\Wi)$. Then the solution is given by Eq. \ref{eq_bzero} (see also Eqs. \ref{eq_blr} and \ref{eq_pp}), with $Y$ replaced by $\tilde{Y}$ in Eq. \ref{eq_opt_bzero}. Finally, undoing the substitution, and realizing that the diagonal matrix  $\dMat(\Wi)$ can be pulled out of both terms in Eq. \ref{eq_bzero}, we obtain
\begin{equation}
\Bs = \Bzero \cdot \dMat(\Wi),
\label{eq_rescaled}
\end{equation}
where $\Bzero$ is given by Eq. \ref{eq_bzero}.
This solution has the interesting property that it decomposes such that the re-scaling with $\Wi$ can be applied \emph{after} $\Bzero$ has been learned. This is especially useful in practical situations where the weights $\Wi$ may change rapidly: the model does not need to be re-trained with new weights $\Wi$--instead one may simply multiply the learned model $\Bzero$  with the current weights $\Wi$ at the time of making recommendations.

Moreover, note that the item-weights  $\Wi$ are unary quantities, while the model-weights $B$ are pairwise quantities--hence, compared to $B$, the item-weights $\Wi$ can be estimated reliably from a much smaller data set, which is beneficial in case of rapid changes in item-popularities. 
\subsection{Example: Popularity Adjustments}
\label{sec_pop}
When  training without re-weighting or re-scaling, collaborative-filtering approaches learn not only the similarities among the items but also their different popularities from the training data. This learned trade-off between item-similarities and item-popularities determines the recommendations made for a given user. Adjusting this trade-off can be crucial for  the quality of  recommendations \cite{castells18,steck11}. 
We illustrate the effectiveness of the re-scaled approach (see Eqs. \ref{eq_opt_rescaled} and \ref{eq_rescaled}) in two applications. 
\subsubsection{Removal of Popularity-Bias}
The goal of removing the popularity bias that is present in the training data, is to learn a model that focuses on item \emph{similarities}. We start by 
 defining the
popularity of item $i$ as $\pop_i = \sum_u Y_{u,i}$, and the popularity-vector regarding all items as $\pop = (\pop_1,...,\pop_{|\I|})^\top$. For instance, if $Y$ is  binary, it is the number of users who interacted with item $i$.
As to allow for different degrees of item-popularities to be removed during training, we introduce the exponent $\alpha\in [0,1]$, see also \cite{steck11, mikolov13}. The value of $\alpha$ has to be chosen depending on the data (e.g., 0.75 is used in word2vec for natural language \cite{mikolov13}). We found  $\alpha=0.5$ to work well on the publicly available data used in our experiments in Section \ref{sec_exp}.  Analogous to inverse propensity weighting,  we  chose the weights for re-scaling as
$$\Wi \propto \vec{1} \oslash \pop^\alpha,$$
where $\oslash$ denotes the elementwise division, and the exponent $\alpha$ is applied elementwise as well. Note that the normalization of the weight-vector $\Wi$ does not matter in Eq. \ref{eq_rescaled} when ranking the items according to the predicted scores.

{\bf Experiments:}  Table \ref{tab_thematrix} illustrates the effect: as an example, we picked the movie 'The Matrix (1999)' in the \emph{Netflix} data:$^{\rm 3}$  for a dummy-user who has watched only this movie,  the top recommendations based on $\Bs$ are comprised of three additional sequels, as expected for a model that focuses on  similarity. This is in contrast to the unweighted model $\Bzero$, whose recommendations  reflect the trade-off between item-similarities and the item-popularities as learned from  the training data, which results in only one sequel in the top recommendations.
\subsubsection{Adjusting to Current Item-Popularities}
\label{sec_poptime}
In this section, we show that the re-scaling approach in Eqs. \ref{eq_opt_rescaled} and \ref{eq_rescaled} is a simple yet effective method for adapting the recommendations to the varying item-popularities over time.\footnote{If additionally the user-preferences change over time, or if the sequential aspect of the user-item interactions is crucial,  more powerful models are required, e.g., \cite{tang17,liu18,hidasi17}.}
When making recommendations at time $t$, it is typically suboptimal to use the average popularity $\pop_i$ of an item in the training data, as these data may have been collected over an extended time-period, during which each item's popularity may have changed. Instead, using the items' popularities at time $t$ can lead to improved recommendations. These popularities may be estimated based on a small time-window near $t$, and we denote them  by $\pop_i(t)$. The weights  for re-scaling  can now be defined as 
$$\Wi(t) \propto \pop(t)^\alpha \oslash \pop^\alpha,$$
i.e., in Eq. \ref{eq_rescaled} this removes the  average item-popularity $\pop$ that was learned by $\Bzero$ and replaces it by the popularity $\pop(t)$ at time $t$. As a result, $\Bs$ reflects the item-popularities at time $t$, besides the item-item similarities learned by  $\Bzero$, see Eq. \ref{eq_rescaled}. 

{\bf Experiments:} 
For a dummy-user who  interacted only with the movie 'The Matrix (1999)',
Table \ref{tab_thematrix} (bottom) shows the top recommendations  at three different points in time during the 6-year time-span covered by the \emph{Netflix} data.\footnote{We used the \emph{Netflix} data here, as it provided the date of the user-item interaction, which was unavailable in the \emph{MSD} data.}
 As expected, the sequels make their appearances in the top recommendations according to the rise and fall of their popularities over time. This illustrates the importance of accounting for the item-popularities at the time of making recommendations.

This is quantified in rows 5-10 in Table \ref{tab_nflx},  
which  illustrates the improvements in ranking accuracy when taking into account  the time of recommendation: first, we estimated the item-popularities $\pop(t)$, and hence weights $\Wi(t)$,  by splitting the data of the training-users (which are disjoint from the test-users) into $N$ successive time-intervals with equal amounts of data (the time-intervals may hence have different lengths). Then,
 for each interaction of a test-user with a test-item, we  determined as to which of the $N$ intervals it fell into, and applied the corresponding weights $\Wi(t)$ according to Eq. \ref{eq_rescaled}. This resulted in the ranked list of recommendations, and we determined the rank of the test-item. We finally used these ranks of the test-items as to compute the ranking metrics in the same way as we did for the other, time-agnostic, approaches in Table \ref{tab_nflx}. 
Note that we chose this scheme as it follows exactly the same evaluation-protocol, and uses exactly the same training and test data as was used for the other, time-agnostic, approaches in Table \ref{tab_nflx}. It hence directly shows the large improvements due to taking into account the time of prediction. This evaluation-scheme, of course, is unrealistic, given that  information in the training-data from the future is possibly used when making recommendations for the test-users in the past--which is a general shortcoming of \emph{randomly} splitting the data into test and training sets, as is commonly done in the literature. Predicting the future based on the past, relative to a chosen reference-point in time, of course, would be more realistic.

Rows 5-10  in Table \ref{tab_nflx} also show that splitting the training data into about  $200$ time-intervals yielded the best ranking accuracy. This implies that a smaller number of time-intervals did not provide sufficient resolution in time, while a larger number  reduced the amount of data in each interval to a level where the weights $\Wi$ could not be estimated accurately. 

\section{Sparse Modeling}
\label{sec_sparse}
In this section, we focus on learning a \emph{sparse} weight-matrix $\Bsparse$, instead of a dense one. For simplicity of argument, we assume that $X=Y$ throughout this section: in this case, the weight-matrix $\BzeroXX$ is completely determined by $\PP$, see Eq. \ref{eq_bzeroxx}. If we further assume that the data-matrix $\frac{1}{|\UU|}\cdot X^\top X$ is the covariance matrix (i.e., the means have been subtracted from each column in $X$, see also footnote 1), then the problem of determining $\PP$ (see Eq. \ref{eq_pp}) becomes the estimation problem of a sparse inverse covariance matrix, or sparse precision matrix. This is a well-studied problem in the areas of structure-learning in  graphical models and of statistical signal processing, e.g., \cite{meinshausen06,friedman07,banerjee08,schmidtthesis,guillot12,wiesel12,hsieh13,treister14}. Most of the common approaches add a sparsity-inducing penalty term to the training objective. 
L$_0$-norm  penalties (like AIC or BIC \cite{akaike73,akaike74,schwarz78}) lead to non-convex (and NP hard) optimization problems, which are typically tackled by various (heuristic) subset selection methods, while the  L$_1$-norm  penalty results in a convex optimization problem, which has attracted much recent work, e.g., see \cite{meinshausen06,friedman07,banerjee08,schmidtthesis,wiesel12,hsieh13,treister14} and references therein.

In Section \ref{sec_sparse_rel}, we  point out  that sparse modeling may improve the quality of recommendations by reducing the number of trust-busters in the sense of eliminating generally popular items that are unrelated to the user's interests. As there are typically only few trust-busters, we will discuss in Section \ref{sec_sparse_acc} as to why sparse modeling in collaborative-filtering may not lead to notable improvements in (aggregate) ranking metrics. Finally,  in Section \ref{sec_sparse_approx} we outline a simple \emph{approximate} approach to  learning a sparse weight-matrix  in a computationally efficient way.
\subsection{Relevance of  Non-Zero Weights}
\label{sec_sparse_rel}
In this section, we discuss that
\begin {enumerate}

\item the learned sparsity pattern in $\PP$ (i.e., \emph{which} entries are non-zero) is determined by the item-item similarities irrespective of the item-popularities, hence focusing on relevance (i.e., statistical dependence). 

\item In contrast, the 
 non-zero  \emph{values} in $\PP$ capture both item-item similarities \emph{and} item-popularities when $X$ reflects  sparse binary user-item interactions\footnote{In this case, each column may be viewed approximately as a sample from a Poisson distribution with mean $\mu_i={\rm pop}_i / |\UU|$ and standard deviation $\sqrt{\mu_i}$, where ${\rm pop}_i$ is the number of users who interacted with item $i$.} (where the means are  possibly subtracted from the columns).
\end {enumerate}
Item (2) is immediately evident from the fact that $\PP$ is the (regularized) inverse of $X^\top X$, see Eq. \ref{eq_pp}. Note that the item-popularities  not only affect the mean, but also the standard deviation in  each column of $X$.$^{\rm 4}$ The latter is preserved  even  if $\frac{1}{|\UU|}\cdot X^\top X$ is the covariance matrix.

Item (1) follows from the well-known fact that
 a zero entry $\PP_{i,j}=0$ in the precision matrix (or inverse covariance matrix) corresponds to the conditional independence of the variables $i,j$ given all the other variables $\I\setminus \{i,j\}$ (e.g., \cite{meinshausen06}). It is important to realize that the degree of (conditional) dependence of two Gaussian random variables is unaffected by their means and standard deviations--for this reason,  the various hypothesis-tests regarding the (conditional) independence of Gaussian random variables are based on their (partial) \emph{correlation coefficients}, rather than on their covariances. Note that  the mean and the standard deviation (which  contain information on the item-popularities) do not affect the correlation coefficients.

Given that the learned weight matrix $\BzeroXX$ is completely determined by $\PP$ if we assume $X=Y$ (see Eq. \ref{eq_bzeroxx}), items (1) and (2) hence carry over to  $\Bsparse$. In practice, this may reduce the risk of recommending items that are generally popular but unrelated to a user's past user-item-interactions, as the corresponding entries in $\Bsparse$ are likely learned to be zero for unrelated but popular items due to small \emph{correlations} (while the covariances might possibly be large).

{\bf Experiments:} Our experiments on sparsity are based on \emph{MSD}, as it is the largest of the three data sets, and hence the largest speed-up in training-time can be expected (see Section \ref{sec_complexity}), which is the main goal of using sparse modeling in this paper.  Rows 5-10  in Table \ref{tab_msd} illustrate, for two sparsity levels (0.003 and 0.0007), that the sparsity pattern is indeed determined in good approximation by the correlation matrix: for simplicity,  we created (suboptimal) sparse weight-matrices $\Bsparse$ by elementwise multiplication of the dense solution $\BzeroXX$ with various sparse binary indicator-matrices $A$, which were determined by thresholding the absolute values in three different matrices: (1) thresholding $\BzeroXX$ serves as a baseline, and may also be viewed as subset selection based on an L$_0$-norm penalty. (2) thresholding the correlation matrix ${\rm cor}(X,X)$ is only slightly worse in Table \ref{tab_msd}--even though only (marginal) correlations are considered here. From a computational perspective, this thresholding  has the advantage that it can be carried out \emph{before} learning the weight matrix, which will be used in the algorithm outlined in Section \ref{sec_sparse_approx}. (3) In contrast, thresholding $X^\top X$ yields considerably worse results in Table \ref{tab_msd}, as expected.
Apart from that, it is remarkable that, relative to the dense solution (cf. $\BzeroXX$ in row 3),  the  sparse solutions resulted in  ranking-accuracies that were only slightly degraded at the sparsity levels 0.003 and 0.0007--dense models with the same number of parameters would be restricted to only 123 and 27 latent dimensions and one hidden layer, respectively. The models based on low-dimensional embeddings in rows (b)-(e) in  Table \ref{tab_msd}, however, have a much larger number of parameters--yet their ranking accuracies are considerably worse. 
This illustrates the effectiveness of high-dimensional sparse models compared to deep low-dimensional dense models in this domain.
\subsection{Predictive Accuracy}
\label{sec_sparse_acc}
Given that sparsity in the model-parameters entails regularization of the learned model as well as feature selection,  improved predictive accuracy of sparse models has been observed in various fields, especially when the training data were small, like in bio-informatics. 

In our experiments, however, we did not observe a large difference between dense and sparse solutions: cf. row (4) with row (a)  in  Tables \ref{tab_ml20m} and \ref{tab_nflx}:  '$\BzeroXXno$ $\ge 0$' is the (suboptimal) non-negative solution,  obtained by setting all the negative values in $\BzeroXXno$ to zero (about 60\% of the entries); '$\BzeroXXno$ $\ge 0$' hence is a dense matrix regarding the remaining 40\% of positive entries. The only difference between row (4) and row (a) is hence that \slim{} \cite{ning11}  is   additionally  a sparse model.\footnote{\slim{} was trained with the original code published by the authors of \cite{ning11}, and hence is a close-to-optimal solution. 
} 

While this empirical result of about equal predictive accuracy of sparse and dense models in our experiments may by surprising at first glance, it may also be explained as follows:
predictive accuracy is typically evaluated in terms of cross validation (or held-out test-data). Now, let us recall two properties of AIC \cite{akaike73,akaike74}: (1) AIC is obtained as the leading-order approximation to cross-validation in the asymptotic limit \cite{akaike73,akaike74}; (2) when AIC is added as a penalty-term to the training objective, while it may entail sparse solutions when the training set is small, AIC tends to entail (close to) dense solutions for large data sets, as the (unknown) true model underlying the data is typically outside the (limited) model-class considered.  In our experiments, the amount of data is apparently sufficiently close to the asymptotic limit
 (in aggregate across all users, which determines the sufficient statistics for training, i.e., the data matrices $X^\top X$ and $X^\top Y$), so that sparse models may not achieve considerably improved prediction/ranking accuracy compared to dense models.
\subsection{Efficient Approximate Sparse Training}
\label{sec_sparse_approx}
While the computational cost (memory footprint and computation time) can be greatly reduced in sparse models  when  making predictions/recommendations,  \emph{learning}  a sparse model often has a larger computational cost than learning a dense model (see also introduction to Section \ref{sec_sparse}). 
For this reason, we now outline a simple heuristic for obtaining a sparse solution in a computationally efficient way, comprised of three steps.

First, we determine the sparsity-pattern $A\in \{0,1\}^{|\I|\times |\I|}$ of $\Bsparse$ by applying a threshold $\theta$ to the absolute value of the (marginal) correlation coefficients (see also Section \ref{sec_sparse_rel}): $A_{i,j}=0$ if $|{\rm cor}(X,X)| <\theta$, and $A_{i,j}=1$ otherwise. The value of $\theta$ may be chosen according to the desired p-value in the hypothesis test for independence of Gaussian variables, according to the corresponding L$_0$-norm penalty-term added to the training objective,  or simply such that the desired level of sparsity is obtained.

Note that, under the Markov assumption in Markov networks (but not in Bayesian networks), it holds that   conditional independence of $i$ and $j$ given a set ${\mathcal{S}} \subseteq \I\setminus\{i,j\}$ implies that they are also independent conditional on any super-set ${\mathcal{T}}$ of ${\mathcal{S}}$. Under this assumption, the marginal independence of $i$ and $j$ (as determined by threshold $\theta$) implies that $\PP_{ij}=0$.

This first step may be viewed as a backward subset-selection step or as the initial step of the constraint-based approach to learning graphical models \cite{spirtes93}--for computational efficiency we do not consider (higher-order) partial correlations here. The goal of the first step merely is to efficiently determine a sparsity pattern such that the second step can be computed efficiently (where additional (close to) zero entries in  $\PP$ may be determined). To this end, we additionally cap the number of non-zero entries in each column of $A$ by $N^{\rm (max)}$ (we chose 1,000 in our experiments), which limits the maximal size of the sub-problems to be solved in the second step.

In the second step,  we estimate the non-zero values in  $\Bsparse$ given the sparsity pattern in $A$ from the first step. The non-zero values may be computed exactly by solving a separate regression problem for each column of $\Bsparse$, as was done in fsSLIM \cite{ning11}. Following the theme of this paper, we instead aim to solve the regression problem for an entire sub-matrix (i.e., \emph{several} columns) at once for computational efficiency. To this end, we start by maintaining a list $\LL$ of the column-indices of $A$,  sorted in descending order by the number of non-zero entries in each column of $A$ (as a tie-break, we use the maximal correlation coefficient $|{\rm cor}(X,X)|$ (i.e., absolute value) in each column as a secondary sorting criteria). 

We then iterate through  the list $\LL$  until it is empty as follows: at step $k$ of the iteration, if $i$ is the first column-index in list $\LL$, we  determine the set $ \I_k$ of item-indices $j$ where $A_{j,i}=1$. We then remove all the indices  $j \in \I_k$ from list $\LL$. Note that $\LL$ shrinks in size by several indices per iteration, which makes this approach computationally efficient.  Now, we estimate the sub-matrix $\Bsparse_{\I_k \times \I_k}$ from the (dense) sub-matrix $(X^\top X)_{\I_k \times \I_k}$ according to Eqs. \ref{eq_bzeroxx} and \ref{eq_pp}.

Estimating these sub-matrices independently of each other in each step, may admittedly  be a crude approximation in general. If   $A$ is a block-diagonal matrix, however, the exact solution is obtained.
Given that each sub-matrix$(X^\top X)_{\I_k \times \I_k}$  is concerned with a set $\I_k$ of highly-correlated items by construction (see first step), matrix $A$ may actually be close to block-diagonal in some sense, with some overlap of the blocks.

In the third and final step, all the sub-matrices $\Bsparse_{\I_k \times \I_k}$ are aggregated as to obtain $\Bsparse$, by simply averaging their values where these sub-matrices overlap.

{\bf Experiments:}  Regarding \emph{MSD}, the largest data set in our experiments,
the experimental results are shown in rows 11-12 in Table \ref{tab_msd}: the ranking accuracies drop only slightly compared to the dense solution (row 2), while still considerably outperforming the low-dimensional-embedding models in rows (b)-(e) in Table \ref{tab_msd}. At the same time the training-time is greatly reduced, as discussed in the next section.

\section{Computational Cost}
\label{sec_complexity}
The computational cost of the presented approach is determined by the size of the matrices $X^\top X$ and $X^\top Y$, which  can serve as sufficient statistics in place of the possibly much larger matrices $X$ and $Y$. They can be computed in a pre-processing step prior to learning the model. The step that is computationally expensive is the matrix inversion to obtain $\PP$, see Eq. \ref{eq_pp}.   The  computational complexity of a matrix inversion is about $\OO(|\I|^{2.376})$ when using the Coppersmith-Winograd algorithm.

The closed-form solution was key to the vastly reduced training-times   in our experiments: learning $\BzeroXXno$ took less than 2, 2 and 20 minutes on the data sets \emph{ML20M},  \emph{Netflix} and \emph{MSD}, respectively,  on an AWS instance with 64 GB RAM and 16 vCPUs. In stark contrast,  \cite{liang18} reports that parallelized grid search for \slim{} took about two weeks on the \emph{Netflix} data, and the \emph{MSD} data was 'too large for it to finish in a reasonable amount of time' \cite{liang18}. Apart from that, the variational autoencoders, the most accurate models among the baselines, took several hours to train, using the publicly available code.$^{\rm 6}$

When learning the sparse approximation $\Bsparse$ on the \emph{MSD} data (see Table \ref{tab_msd}), we observed that
 the (wall-clock) training-time  dropped from less than 20 minutes for (dense) $\Bzero$ (row 3) to less than 2 minutes (row 11) and 30 seconds (row 12) for the sparse approximation. Moreover,  this sparse approximation also reduces the memory footprint during training (steps 2 and 3), as only small sub-matrices have to be kept in memory. Step 1 is memory-intense, but requires only  simple thresholding-operations that can be implemented on any common big-data platform for pre-processing the data.
\section{Experiments}
\label{sec_exp}
In this section, we summarize the experimental set-up. 
We follow  the setting in \cite{liang18}, as the authors provided  publicly available code$^{\rm 6}$ for reproducibility of the results.
 Our experimental results are discussed in the corresponding previous sections regarding zero-diagonals, biased data and sparse modeling.

\begin{table}[t]
\caption{Accuracy on \emph{ML-20M} data (standard errors$\approx$0.002).}
\label{tab_ml20m}
\begin{tabular}{rlrrr}
row & approach  &   Recall@20    &    Recall@50    &    NDCG@100\\
\toprule
(1)   &  popularity    &    0.162    &    0.235    &   0.191 \\
\midrule
(2)   &  $\Blr$     &    0.375    &    0.507    &   0.406 \\
(3)   &  $\BzeroXXno$     &    0.391    &    0.521    &   0.420 \\
(4)   &  $\BzeroXXno$  $\ge 0$    &  0.373    &  0.499  &  0.402\\
\midrule
\multicolumn{5}{l}{ results reproduced from \cite{liang18}:}\\

(a)   &  \slim{}    &    0.370    &    0.495    &    0.401\\
(b)   &  \wmf{}     &    0.360    &    0.498    &    0.386\\
(c)   &  \cdae{}     &    0.391    &    0.523    &    0.418\\
(d)   &  \mvae{}     &    0.395    &    0.537    &    0.426\\
(e)   &  \mdae{}     &    0.387    &    0.524    &    0.419\\
\bottomrule
\end{tabular}

\end{table}

\begin{table}[t]
\caption{\emph{Netflix} data:  Top recommendations for a dummy-user who watched only the  'The Matrix (1999)', using different weighting schemes, see Section \ref{sec_pop}.}
\label{tab_thematrix}
\begin{tabular}{l}
\toprule
{\bf $\Bzero$ : unweighted}\\
 The Matrix: Reloaded (2003)\\
 Gladiator (2000)\\
 Men in Black (1997)\\
 Fight Club (1999)\\
 Lord of the Rings: The Fellowship of the Ring (2001)\\
 Minority Report (2002)\\

\midrule
{\bf $\Bs$ : movie-popularities removed:}\\
 The Matrix: Reloaded (2003)\\
 Gladiator (2000)\\
 The Matrix: Revolutions (2003)\\
 The Fifth Element (1997)\\
 Men in Black (1997)\\
 The Matrix: Revisited (2001)\\
\midrule

{\bf $\Bs$ :  movie-popularities adjusted over time:}\\
{\bf  time-interval (1999-11-11 ... 2000-09-02):}\\
 The Fifth Element (1997)\\
 The Terminator (1984)\\
 The Sixth Sense (1999)\\
 Saving Private Ryan (1998)\\
 The Silence of the Lambs (1991)\\
 12 Monkeys (1995)\\
%\hline
{\bf time-interval  (2004-05-29 ... 2004-06-04):}\\ 
 The Matrix: Reloaded (2003)\\
 Gladiator (2000)\\
 The Matrix: Revolutions (2003)\\
 Lord of the Rings: The Fellowship of the Ring (2001)\\
 Fight Club (1999)\\
 Minority Report (2002)\\
%\hline
{\bf  time-interval (2005-12-25 ... 2005-12-31):}\\ 
 The Matrix: Reloaded (2003)\\
 Gladiator (2000)\\
 Men in Black (1997)\\
 Fight Club (1999) \\
 The Fifth Element (1997)\\
 X-Men (2000)\\
\bottomrule
\end{tabular}
\end{table}

Given that a single user-item training-matrix $Z$ was available in the publicly available data sets (instead of two different matrices $X$ and $Y$), we use $X=Y=Z$ for all the models, except for $\Blr$ where we use the modification outlined in Section \ref{sec_xy0d}.

While the reader is referred to \cite{liang18} for details of the experimental setting, we provide a summary in the following.
In \cite{liang18}, results for the following models were reported, which we now use  as baselines in our paper:
\begin{itemize}

\item Sparse Linear Method  (\slim{}) \cite{ning11}. Besides the original model, also a   computationally faster approximation  (which drops the constraints on the weights) \cite{levy13} was considered, but its results were not  found  to be on par with the other models in the experiments in \cite{liang18}.

\item Weighted Matrix Factorization (\wmf) \cite{hu08,pan08}, a linear model with a latent representation of users and items.

\item Collaborative Denoising Autoencoder (\cdae) \cite{wu16}, a non-linear model with one hidden layer.

\item denoising autoencoder (\mdae) and variational autoencoder (\mvae) \cite{liang18}, both trained using the multinomial likelihood, which was found to outperform the Gaussian and logistic likelihoods. Best results were obtained in \cite{liang18} for the \mvae{}  and \mdae{} models that were rather shallow 'deep models', namely with a 200-dimensional latent representation, as well as a 600-dimensional hidden layer in both the encoder and decoder. Both models are non-linear, and \mvae{}  is also probabilistic.

\end{itemize}
Three data sets were used in the experiments in \cite{liang18}, and were pre-processed and filtered for items and users with a certain activity level, resulting in the following data-set sizes, see \cite{liang18} for details:\footnote{The code regarding  \emph{ML-20M} in \cite{liang18} is publicly available at {\tt https://github.com/dawenl/vae\_cf}. Upon request, the authors kindly provided the code for the other two  data sets.}
\begin{itemize}
\item MovieLens 20 Million (\emph{ML-20M}) data \cite{movielens20mio}: 136,677 users and 20,108 movies with about 10 million interactions,
\item Netflix Prize (\emph{Netflix}) data \cite{netflixdata}: 463,435 users and 17,769 movies with about 57 million interactions,
\item Million Song Data (\emph{MSD}) \cite{msddata}: 571,355 users and 41,140 songs with about 34 million interactions.
\end{itemize}

We also follow the evaluation protocol used in \cite{liang18}, which is based on  \emph{strong generalization}, i.e., the training, validation and test sets are disjoint in terms of users. This is in contrast to \emph{weak generalization}, where the training and test sets are disjoint in terms of  user-item interaction-pairs, but not in terms of users. 
Concerning  evaluation in terms of ranking metrics, Recall@$k$ for $k\in\{20,50\}$ as well as Normalized Discounted Cumulative Gain, NDCG@100 were used  in \cite{liang18}.

When learning $\Bzero$, we found  the optimal L2-norm regularization parameter $\lambda$ to be about 500 on \emph{ML-20M}, 1,000 on \emph{Netflix}, and 200 on \emph{MSD} data. Note that these values are much larger than the typical values used for \slim{}, which often are of the order of 1, see \cite{ning11}. $\Bzero$ is dense and hence has many more parameters than than \slim{}, which is sparse. In the sparse approximation outlined in Section \ref{sec_sparse_approx}, we found the optimal $\lambda$ to decrease from 200 to 50 and 5 for sparsity levels 0.003 and 0.0007, respectively, on the \emph{MSD} data (see rows 11 and 12 in Table \ref{tab_msd}).

As mentioned earlier, the experimental results regarding the different variants of linear regression are discussed in the corresponding Sections \ref{sec_0d}, \ref{sec_covshift}, and \ref{sec_sparse} above.

\begin{table}[t]
\caption{Accuracy on \emph{Netflix}  data (standard errors$\approx$0.001).}
\label{tab_nflx}
\begin{tabular}{rlrrr}
row &approach    &   Recall@20    &    Recall@50    &    NDCG@100\\
\toprule
(1) &  popularity    &    0.116    &    0.175    &   0.159 \\
\midrule

(2) &  $\Blr$     &    0.349    &    0.434    &   0.380 \\
(3)&  $\BzeroXXno$     &    0.362    &    0.445    &   0.393 \\
(4)&  $\BzeroXXno$  $\ge 0$    &  0.345    &  0.424  &  0.373\\
\midrule
\multicolumn{4}{l}{$\Bs$: time intervals  (Section \ref{sec_poptime}):}\\
(5)&   5              &  0.392 & 0.471     &  0.422    \\
(6)&   10            & 0.407  &  0.482    &  0.436    \\
(7)&   50            & 0.426  &  0.494    &  0.455    \\
(8)&   100           & 0.430  &  0.497    &  0.459    \\
(9)&   200           & 0.432  &  0.498    &  0.461    \\
(10)&   500          & 0.425  &   0.490   &   0.453   \\
\midrule
\multicolumn{4}{l}{ results reproduced from \cite{liang18}:}\\

(a) &  \slim{}    &    0.347    &    0.428    &    0.379\\
(b) &  \wmf{}     &    0.316    &    0.404    &    0.351\\
(c) &  \cdae{}     &    0.343    &    0.428    &    0.376\\
(d) &  \mvae{}     &    0.351    &    0.444    &    0.386\\
(e) &  \mdae{}     &    0.344    &    0.438    &    0.380\\
\bottomrule
\end{tabular}

\end{table}

{\bf Difference between Data Sets:}
When comparing the ranking accuracies across the three data sets (see  Tables \ref{tab_ml20m}, \ref{tab_nflx}, and \ref{tab_msd}), it is interesting that, relative to the best competing model,  $\BzeroXXno$ is slightly worse on \emph{ML-20M}, slightly better on \emph{Netflix}, and considerably better on \emph{MSD} (remarkably by about 20\%). Having considered various properties of these data sets (see also table 1 in \cite{liang18}), we suspect that this may be explained by the  trade-off between recommending generally popular items vs. personally relevant items to each user: to this end, we evaluated  the popularity-based model (see row 1  in Tables \ref{tab_ml20m}, \ref{tab_nflx}, and \ref{tab_msd}) as an additional baseline,  i.e.,  the items are ranked by their popularities. These unpersonalized recommendations obviously ignore the personalized relevance to a user. 
 Row 1 in Tables \ref{tab_ml20m}, \ref{tab_nflx}, and \ref{tab_msd} shows that this  popularity-based model obtains better accuracy on the \emph{ML-20M} data than it does on the \emph{Netflix} data, while its accuracy is considerably reduced on the \emph{MSD} data. This suggests that good recommendations on the  \emph{MSD} data have to focus much more on personally relevant items rather than on generally popular items, compared to the  \emph{ML-20M} and \emph{Netflix} data. The notable  improvement of $\BzeroXXno$ over the competing  models on the   \emph{MSD} data suggests that it is able to better recommend personally relevant items on this data set.
 On the other hand, the results on  the  \emph{ML-20M} and \emph{Netflix} data suggest that  $\BzeroXXno$ is also able to make recommendations with an increased focus on generally  popular items if necessary.

\begin{table}[t]
\caption{Accuracy on \emph{MSD}  data (standard errors$\approx$0.001).}
\label{tab_msd}
\begin{tabular}{rllll}
  &   &   Recall    &    Recall    &    NDCG \\
 
  row           &  approach      &   @20    &    @50   &    @100 \\
  \toprule
 
(1)  &  popularity    &    0.043    &    0.068    &   0.058 \\
\midrule

(2)  &  $\Blr$     &    0.324    &    0.422    &   0.379 \\
(3)  &  $\BzeroXXno$     &    0.333    &    0.428    &   0.389 \\
(4)  &  $\BzeroXXno$  $\ge 0$    &  0.324    &  0.418  &  0.379\\
\midrule
\multicolumn{5}{l}{sparse approximation (Sec. \ref{sec_sparse_rel}): $A \odot \BzeroXXno$ for various $A$: }\\
&\multicolumn{4}{l}{sparsity level 0.003: }\\
(5)  & $|\BzeroXXno | \ge 0.0075   $   & 0.331   & 0.425      & 0.387  \\
(6)  & $|\cor(X^\top X)| \ge 0.03   $   & 0.331   & 0.424      & 0.387  \\
(7)  & $X^\top X \ge 50$     & 0.327   & 0.418      & 0.381  \\
&\multicolumn{4}{l}{sparsity level 0.0007: }\\
(8)  & $|\BzeroXXno | \ge 0.013   $   & 0.329   & 0.420      & 0.384  \\
(9)  & $ |\cor(X^\top X)| \ge 0.1   $   & 0.324   & 0.412      & 0.377  \\
(10)  & $X^\top X \ge 140$  & 0.292   & 0.367      & 0.342  \\

\midrule
\multicolumn{5}{l}{$\Bsparse$: sparse block-wise  approximation (Section \ref{sec_sparse_approx}):}\\
(11)  & sparsity level 0.003    & 0.326   & 0.419      & 0.380  \\
(12)  & sparsity level 0.0007  &  0.319  &  0.405     & 0.371  \\
\midrule
\multicolumn{5}{l}{ results reproduced from \cite{liang18}: }\\

(a)  & \slim{}    &   \multicolumn{3}{c}{ --- did not finish in \cite{liang18} ---  }\\
(b)  & \wmf{}     &    0.211    &    0.312    &    0.257\\
(c)  & \cdae{}     &    0.188    &    0.283    &    0.237\\
(d)  & \mvae{}     &    0.266    &    0.364    &    0.316\\
(e)  & \mdae{}     &    0.266    &    0.363    &    0.313\\
\bottomrule
\end{tabular}
\end{table}

\section{Conclusions and Future Work}

As the contributions of this paper are itemized in the Introduction, we conclude with a research question raised by the empirical results in our experiments: with similar memory footprints, will deep models that use \emph{high-dimensional sparse} representations be  considerably more accurate than deep models based on \emph{low-dimensional dense} embeddings, especially in domains with a large number of diverse items? This question is motivated by the
observation that
the (shallow) linear regression models considerably outperformed all the competing deep low-dimensional models on the task of making   highly personalized recommendations (see Million Song Data in Table \ref{tab_msd}). 
 The linear model is based on the full-rank item-item matrix, and hence is  high-dimensional. 
In large domains,  a  high-dimensional dense model may not fit into memory, which calls for sparse versions of  high-dimensional models--both variants obtained similar accuracies in our experiments, even when very sparse. It will be interesting to see if 'going deep' with high-dimensional sparse models will lead to  similar gains in accuracy as were observed when 'going deep' with  models based on low-dimensional dense  embeddings.
\section{Appendix}
If a single binary user-item interaction-matrix $Z\in\{0,1\}^{|\UU| \times |\I|}$ is  available for training, we generate the  disjoint training-matrices  $Y,X$ needed for $\Blr$ in Eq. \ref{eq_blr} as follows: we split the observed user-item interactions  in $Z$ into two disjoint sets, one assigned to Y and one to X. Let us assume that the split is done randomly, and a fraction $p$ of a user's interactions is assigned to $Y$, and the remaining fraction $1-p$ to $X$. Instead of using a particular split, we use the expectation over the various splits: it is easy to see for binary $Z$ that
\begin{eqnarray*}
&& \EE \left [X^\top Y \right] = \sum_{u \in \UU} \EE \left[ X_{u,\cdot}^\top Y_{u,\cdot} \right ]\\
&&\,\,=   \sum_{u \in \UU} p(1-p)\left( Z_{u,\cdot}^\top Z_{u,\cdot} - \dMat(\diag( Z_{u,\cdot}^\top Z_{u,\cdot})) \right )\\
&&\,\,=  p(1-p)\left( Z^\top Z - \dMat(\diag( Z^\top Z))  \right )\\
&&\,\,\propto  Z^\top Z - \dMat(\diag( Z^\top Z))  ,
\end{eqnarray*}
where the diagonal is zero, as expected for disjoint $X,Y$, cf. Section \ref{sec_xy0d}, and the off-diagonal values are proportional to the ones in $Z^\top Z$.
Moreover, 
\begin{eqnarray*}
&& \EE \left [X^\top X \right] = \sum_{u \in \UU} \EE \left[ X_{u,\cdot}^\top X_{u,\cdot} \right ]\\
&&\,\,=   \sum_{u \in \UU} (1-p)^2\left( Z_{u,\cdot}^\top Z_{u,\cdot} \right )\\
&&\,\,\,\,\,\,\,\,\,\,\,\,\,\,\,\,\,\,\,\,\,\,\,\,\,\,
+ \left( (1-p)-(1-p)^2  \right) \cdot \dMat(\diag( Z_{u,\cdot}^\top Z_{u,\cdot}))\\
&&\,\,=  (1-p)^2 Z^\top Z + p (1-p) \cdot \dMat(\diag( Z^\top Z))\\
&&\,\,\approx  (1-p)^2 Z^\top Z\,\,\,\,\,{\rm for\,\,\, small}\,\,\,p\\
&&\,\,\propto   Z^\top Z  .
\end{eqnarray*}
Note that the diagonal values are increased relative to the off-diagonal ones.  In  Eq. \ref{eq_pp}, this implicitly causes an additional L$_2$-norm regularization of $\PP$, similar to $\lambda$. As to explicitly control for the   L$_2$-norm regularization via the parameter $\lambda$ in our experiments, we use the approximation that is valid for a very small value $p$. Finally, we drop the irrelevant proportionality constants.

\begin{acks}
 I am very grateful to Tony Jebara, Maria  Dimakopoulou, and Nickolai Riabov for useful comments on an earlier draft. I am especially thankful to  Dawen Liang for providing the code used for his paper and for numerous insightful discussions.
\end{acks}

\bibliographystyle{ACM-Reference-Format}

\begin{thebibliography}{37}

%%% the Web address.

\ifx \showCODEN    \undefined \def \showCODEN     #1{\unskip}     \fi
\ifx \showDOI      \undefined \def \showDOI       #1{#1}\fi
\ifx \showISBNx    \undefined \def \showISBNx     #1{\unskip}     \fi
\ifx \showISBNxiii \undefined \def \showISBNxiii  #1{\unskip}     \fi
\ifx \showISSN     \undefined \def \showISSN      #1{\unskip}     \fi
\ifx \showLCCN     \undefined \def \showLCCN      #1{\unskip}     \fi
\ifx \shownote     \undefined \def \shownote      #1{#1}          \fi
\ifx \showarticletitle \undefined \def \showarticletitle #1{#1}   \fi
\ifx \showURL      \undefined \def \showURL       {\relax}        \fi
% invisible to TeX
\providecommand\bibfield[2]{#2}
\providecommand\bibinfo[2]{#2}
\providecommand\natexlab[1]{#1}
\providecommand\showeprint[2][]{arXiv:#2}

\bibitem[\protect\citeauthoryear{Akaike}{Akaike}{1973}]%
        {akaike73}
\bibfield{author}{\bibinfo{person}{H. Akaike}.}
  \bibinfo{year}{1973}\natexlab{}.
\newblock \showarticletitle{Information Theory and an extension of the maximum
  likelihood principle}. In \bibinfo{booktitle}{\emph{Second International
  Symposium on Information Theory}}, \bibfield{editor}{\bibinfo{person}{B.~N.
  Petrox} {and} \bibinfo{person}{F.~Caski}} (Eds.). \bibinfo{address}{Akademia
  Kiado, Budapest}, \bibinfo{pages}{267--81}.
\newblock


\bibitem[\protect\citeauthoryear{Akaike}{Akaike}{1974}]%
        {akaike74}
\bibfield{author}{\bibinfo{person}{H. Akaike}.}
  \bibinfo{year}{1974}\natexlab{}.
\newblock \showarticletitle{A new look at the Statistical Model
  Identification}.
\newblock \bibinfo{journal}{\emph{IEEE Trans. Automat. Control}}
  \bibinfo{volume}{19} (\bibinfo{year}{1974}), \bibinfo{pages}{716--23}.
\newblock


\bibitem[\protect\citeauthoryear{Banerjee, Ghaoui, and d'Aspremont}{Banerjee
  et~al\mbox{.}}{2008}]%
        {banerjee08}
\bibfield{author}{\bibinfo{person}{O. Banerjee}, \bibinfo{person}{L.E. Ghaoui},
  {and} \bibinfo{person}{A. d'Aspremont}.} \bibinfo{year}{2008}\natexlab{}.
\newblock \showarticletitle{Model Selection Through Sparse Maximum Likelihood
  Estimation for Multivariate Gaussian or Binary Data}.
\newblock \bibinfo{journal}{\emph{Journal of Machine Learning Research}}
  \bibinfo{volume}{9} (\bibinfo{year}{2008}).
\newblock


\bibitem[\protect\citeauthoryear{Bennet and Lanning}{Bennet and
  Lanning}{2007}]%
        {netflixdata}
\bibfield{author}{\bibinfo{person}{J. Bennet} {and} \bibinfo{person}{S.
  Lanning}.} \bibinfo{year}{2007}\natexlab{}.
\newblock \showarticletitle{The {N}etflix {P}rize}. In
  \bibinfo{booktitle}{\emph{Workshop at SIGKDD-07, ACM Conference on Knowledge
  Discovery and Data Mining}}.
\newblock


\bibitem[\protect\citeauthoryear{Bertin-Mahieux, Ellis, Whitman, and
  Lamere}{Bertin-Mahieux et~al\mbox{.}}{2011}]%
        {msddata}
\bibfield{author}{\bibinfo{person}{T. Bertin-Mahieux}, \bibinfo{person}{D.P.W.
  Ellis}, \bibinfo{person}{B. Whitman}, {and} \bibinfo{person}{P. Lamere}.}
  \bibinfo{year}{2011}\natexlab{}.
\newblock \showarticletitle{The Million Song Dataset}. In
  \bibinfo{booktitle}{\emph{International Society for Music Information
  Retrieval Conference (ISMIR)}}.
\newblock


\bibitem[\protect\citeauthoryear{Ca{\~n}amares and Castells}{Ca{\~n}amares and
  Castells}{2018}]%
        {castells18}
\bibfield{author}{\bibinfo{person}{R. Ca{\~n}amares} {and} \bibinfo{person}{P.
  Castells}.} \bibinfo{year}{2018}\natexlab{}.
\newblock \showarticletitle{Should {I} follow the crowd? {A} probabilistic
  analysis of the effectiveness of popularity in recommender systems}. In
  \bibinfo{booktitle}{\emph{ACM Conference on Research and Development in
  Information Retrieval (SIGIR)}}.
\newblock


\bibitem[\protect\citeauthoryear{Friedman, Hastie, and Tibshirani}{Friedman
  et~al\mbox{.}}{2008}]%
        {friedman07}
\bibfield{author}{\bibinfo{person}{J. Friedman}, \bibinfo{person}{T. Hastie},
  {and} \bibinfo{person}{R. Tibshirani}.} \bibinfo{year}{2008}\natexlab{}.
\newblock \showarticletitle{Sparse inverse covariance estimation with the
  graphical lasso}.
\newblock \bibinfo{journal}{\emph{Biostatistics}}  \bibinfo{volume}{9}
  (\bibinfo{year}{2008}).
\newblock
Issue 3.


\bibitem[\protect\citeauthoryear{Guillot, Rajaratnam, Rolfs, Maleki, and
  Wong}{Guillot et~al\mbox{.}}{2012}]%
        {guillot12}
\bibfield{author}{\bibinfo{person}{D. Guillot}, \bibinfo{person}{B.
  Rajaratnam}, \bibinfo{person}{B. Rolfs}, \bibinfo{person}{A. Maleki}, {and}
  \bibinfo{person}{I. Wong}.} \bibinfo{year}{2012}\natexlab{}.
\newblock \showarticletitle{Iterative Thresholding Algorithm for Sparse Inverse
  Covariance Estimation}. In \bibinfo{booktitle}{\emph{Advances in Neural
  Information Processing Systems (NIPS)}}.
\newblock


\bibitem[\protect\citeauthoryear{Harper and Konstan}{Harper and
  Konstan}{2015}]%
        {movielens20mio}
\bibfield{author}{\bibinfo{person}{F.~M. Harper} {and} \bibinfo{person}{J.~A.
  Konstan}.} \bibinfo{year}{2015}\natexlab{}.
\newblock \showarticletitle{The MovieLens Datasets: History and Context}.
\newblock \bibinfo{journal}{\emph{ACM Transactions on Interactive Intelligent
  Systems (TiiS)}}  \bibinfo{volume}{5} (\bibinfo{year}{2015}).
\newblock
Issue 4.


\bibitem[\protect\citeauthoryear{Hidasi and Karatzoglou}{Hidasi and
  Karatzoglou}{2017}]%
        {hidasi17}
\bibfield{author}{\bibinfo{person}{B. Hidasi} {and} \bibinfo{person}{A.
  Karatzoglou}.} \bibinfo{year}{2017}\natexlab{}.
\newblock \showarticletitle{Recurrent Neural Networks with Top-k Gains for
  Session-based Recommendations}. In \bibinfo{booktitle}{\emph{International
  Conference on Information and Knowledge Management (CIKM)}}.
\newblock
\newblock
\shownote{arXiv:1706.03847.}


\bibitem[\protect\citeauthoryear{Hsieh, Sustik, Dhillon, Ravikumar, and
  Poldrack}{Hsieh et~al\mbox{.}}{2013}]%
        {hsieh13}
\bibfield{author}{\bibinfo{person}{C.-J. Hsieh}, \bibinfo{person}{M.A. Sustik},
  \bibinfo{person}{I.S. Dhillon}, \bibinfo{person}{P.K. Ravikumar}, {and}
  \bibinfo{person}{R. Poldrack}.} \bibinfo{year}{2013}\natexlab{}.
\newblock \showarticletitle{Sparse Inverse Covariance Estimation for a Million
  Variables}. In \bibinfo{booktitle}{\emph{Advances in Neural Information
  Processing Systems (NIPS)}}.
\newblock


\bibitem[\protect\citeauthoryear{Hu, Koren, and Volinsky}{Hu
  et~al\mbox{.}}{2008}]%
        {hu08}
\bibfield{author}{\bibinfo{person}{Y. Hu}, \bibinfo{person}{Y. Koren}, {and}
  \bibinfo{person}{C. Volinsky}.} \bibinfo{year}{2008}\natexlab{}.
\newblock \showarticletitle{Collaborative Filtering for Implicit Feedback
  Datasets}. In \bibinfo{booktitle}{\emph{IEEE International Conference on Data
  Mining (ICDM)}}.
\newblock


\bibitem[\protect\citeauthoryear{King and Zeng}{King and Zeng}{2001}]%
        {king01}
\bibfield{author}{\bibinfo{person}{G. King} {and} \bibinfo{person}{L. Zeng}.}
  \bibinfo{year}{2001}\natexlab{}.
\newblock \showarticletitle{Logistic Regression in Rare Events Data}.
\newblock \bibinfo{journal}{\emph{Political Analysis}}  \bibinfo{volume}{9}
  (\bibinfo{year}{2001}).
\newblock


\bibitem[\protect\citeauthoryear{Levy and Jack}{Levy and Jack}{2013}]%
        {levy13}
\bibfield{author}{\bibinfo{person}{M. Levy} {and} \bibinfo{person}{K. Jack}.}
  \bibinfo{year}{2013}\natexlab{}.
\newblock \showarticletitle{Efficient Top-N Recommendation by Linear
  Regression}. In \bibinfo{booktitle}{\emph{RecSys Large Scale Recommender
  Systems Workshop}}.
\newblock


\bibitem[\protect\citeauthoryear{Liang, Charlin, and Blei}{Liang
  et~al\mbox{.}}{2016a}]%
        {liang16b}
\bibfield{author}{\bibinfo{person}{D. Liang}, \bibinfo{person}{L. Charlin},
  {and} \bibinfo{person}{D.M. Blei}.} \bibinfo{year}{2016}\natexlab{a}.
\newblock \showarticletitle{Causal Inference for Recommendation}. In
  \bibinfo{booktitle}{\emph{Causation: Foundation to Application, Workshop at
  UAI}}.
\newblock


\bibitem[\protect\citeauthoryear{Liang, Charlin, McInerney, and Blei}{Liang
  et~al\mbox{.}}{2016b}]%
        {liang16a}
\bibfield{author}{\bibinfo{person}{D. Liang}, \bibinfo{person}{L. Charlin},
  \bibinfo{person}{J. McInerney}, {and} \bibinfo{person}{D.M. Blei}.}
  \bibinfo{year}{2016}\natexlab{b}.
\newblock \showarticletitle{Modeling User Exposure in Recommendation}. In
  \bibinfo{booktitle}{\emph{International World Wide Web Conference (WWW)}}.
\newblock


\bibitem[\protect\citeauthoryear{Liang, Krishnan, Hoffman, and Jebara}{Liang
  et~al\mbox{.}}{2018}]%
        {liang18}
\bibfield{author}{\bibinfo{person}{D. Liang}, \bibinfo{person}{R.~G. Krishnan},
  \bibinfo{person}{M.~D. Hoffman}, {and} \bibinfo{person}{T. Jebara}.}
  \bibinfo{year}{2018}\natexlab{}.
\newblock \showarticletitle{Variational Autoencoders for Collaborative
  Filtering}. In \bibinfo{booktitle}{\emph{International World Wide Web
  Conference (WWW)}}.
\newblock


\bibitem[\protect\citeauthoryear{Liu, Zeng, Mokhosi, and Zhang}{Liu
  et~al\mbox{.}}{2018}]%
        {liu18}
\bibfield{author}{\bibinfo{person}{Q. Liu}, \bibinfo{person}{Y. Zeng},
  \bibinfo{person}{R. Mokhosi}, {and} \bibinfo{person}{H. Zhang}.}
  \bibinfo{year}{2018}\natexlab{}.
\newblock \showarticletitle{{STAMP}: {S}hort-term attention/memory priority
  model for session-based recommendation}. In \bibinfo{booktitle}{\emph{ACM
  Conference on Knowledge Discovery and Data Mining (KDD)}}.
\newblock


\bibitem[\protect\citeauthoryear{Marlin and Zemel}{Marlin and Zemel}{2009}]%
        {marlin09}
\bibfield{author}{\bibinfo{person}{B. Marlin} {and} \bibinfo{person}{R.
  Zemel}.} \bibinfo{year}{2009}\natexlab{}.
\newblock \showarticletitle{Collaborative Prediction and Ranking with
  Non-Random Missing Data}. In \bibinfo{booktitle}{\emph{ACM Conference on
  Recommender Systems (RecSys)}}.
\newblock


\bibitem[\protect\citeauthoryear{Meinshausen and B\"{u}hlmann}{Meinshausen and
  B\"{u}hlmann}{2006}]%
        {meinshausen06}
\bibfield{author}{\bibinfo{person}{N. Meinshausen} {and} \bibinfo{person}{P.
  B\"{u}hlmann}.} \bibinfo{year}{2006}\natexlab{}.
\newblock \showarticletitle{High-dimensional graphs and variable selection with
  the Lasso}.
\newblock \bibinfo{journal}{\emph{Annals of Statistics}}  \bibinfo{volume}{34}
  (\bibinfo{year}{2006}).
\newblock
Issue 3.


\bibitem[\protect\citeauthoryear{Mikolov, Sutskever, Chen, Corrado, and
  Dean}{Mikolov et~al\mbox{.}}{2013}]%
        {mikolov13}
\bibfield{author}{\bibinfo{person}{T. Mikolov}, \bibinfo{person}{I. Sutskever},
  \bibinfo{person}{K. Chen}, \bibinfo{person}{G. Corrado}, {and}
  \bibinfo{person}{J. Dean}.} \bibinfo{year}{2013}\natexlab{}.
\newblock \showarticletitle{Distributed Representations of words and phrases
  and their compositionality}. In \bibinfo{booktitle}{\emph{Conference on
  Neural Information Processing Systems (NIPS)}}.
\newblock


\bibitem[\protect\citeauthoryear{Ning and Karypis}{Ning and Karypis}{2011}]%
        {ning11}
\bibfield{author}{\bibinfo{person}{X. Ning} {and} \bibinfo{person}{G.
  Karypis}.} \bibinfo{year}{2011}\natexlab{}.
\newblock \showarticletitle{SLIM: Sparse Linear Methods for Top-N Recommender
  Systems}. In \bibinfo{booktitle}{\emph{IEEE International Conference on Data
  Mining (ICDM)}}. \bibinfo{pages}{497--506}.
\newblock


\bibitem[\protect\citeauthoryear{Pan, Zhou, Cao, Liu, Lukose, Scholz, and
  Yang}{Pan et~al\mbox{.}}{2008}]%
        {pan08}
\bibfield{author}{\bibinfo{person}{R. Pan}, \bibinfo{person}{Y. Zhou},
  \bibinfo{person}{B. Cao}, \bibinfo{person}{N. Liu}, \bibinfo{person}{R.
  Lukose}, \bibinfo{person}{M. Scholz}, {and} \bibinfo{person}{Q. Yang}.}
  \bibinfo{year}{2008}\natexlab{}.
\newblock \showarticletitle{One-Class Collaborative Filtering}. In
  \bibinfo{booktitle}{\emph{IEEE International Conference on Data Mining
  (ICDM)}}.
\newblock


\bibitem[\protect\citeauthoryear{Schmidt}{Schmidt}{2011}]%
        {schmidtthesis}
\bibfield{author}{\bibinfo{person}{M. Schmidt}.}
  \bibinfo{year}{2011}\natexlab{}.
\newblock \emph{\bibinfo{title}{Graphical Model Structure Learning with
  L1-Regularization}}.
\newblock \bibinfo{thesistype}{Ph.D. Dissertation}. \bibinfo{school}{University
  of British Columbia, Vancouver, Canada}.
\newblock


\bibitem[\protect\citeauthoryear{Schnabel, Swaminathan, Singh, Chandak, and
  Joachims}{Schnabel et~al\mbox{.}}{2016}]%
        {schnabel16a}
\bibfield{author}{\bibinfo{person}{T. Schnabel}, \bibinfo{person}{A.
  Swaminathan}, \bibinfo{person}{A. Singh}, \bibinfo{person}{N. Chandak}, {and}
  \bibinfo{person}{T. Joachims}.} \bibinfo{year}{2016}\natexlab{}.
\newblock \showarticletitle{Recommendations as Treatments: Debiasing Learning
  and Evaluation}. In \bibinfo{booktitle}{\emph{International Conference on
  Machine Learning (ICML)}}.
\newblock


\bibitem[\protect\citeauthoryear{Schwarz}{Schwarz}{1978}]%
        {schwarz78}
\bibfield{author}{\bibinfo{person}{G. Schwarz}.}
  \bibinfo{year}{1978}\natexlab{}.
\newblock \showarticletitle{Estimating the dimension of a model}.
\newblock \bibinfo{journal}{\emph{The Annals of Statistics}}
  \bibinfo{volume}{6}, \bibinfo{number}{2} (\bibinfo{year}{1978}),
  \bibinfo{pages}{461--64}.
\newblock


\bibitem[\protect\citeauthoryear{Sedhain, Menon, Sanner, and Braziunas}{Sedhain
  et~al\mbox{.}}{2016}]%
        {sedhain16}
\bibfield{author}{\bibinfo{person}{S. Sedhain}, \bibinfo{person}{A.~K. Menon},
  \bibinfo{person}{S. Sanner}, {and} \bibinfo{person}{D. Braziunas}.}
  \bibinfo{year}{2016}\natexlab{}.
\newblock \showarticletitle{On the Effectiveness of Linear Models for One-Class
  Collaborative Filtering}.
\newblock \bibinfo{journal}{\emph{AAAI}} (\bibinfo{year}{2016}).
\newblock


\bibitem[\protect\citeauthoryear{Spirtes, Glymour, and Scheines}{Spirtes
  et~al\mbox{.}}{1993}]%
        {spirtes93}
\bibfield{author}{\bibinfo{person}{P. Spirtes}, \bibinfo{person}{C. Glymour},
  {and} \bibinfo{person}{R. Scheines}.} \bibinfo{year}{1993}\natexlab{}.
\newblock \bibinfo{booktitle}{\emph{Causation, Prediction, and Search}}.
\newblock \bibinfo{publisher}{Springer Lecture Notes in Statistics 81}.
\newblock


\bibitem[\protect\citeauthoryear{Steck}{Steck}{2010}]%
        {steck10}
\bibfield{author}{\bibinfo{person}{H. Steck}.} \bibinfo{year}{2010}\natexlab{}.
\newblock \showarticletitle{Training and Testing of Recommender Systems on Data
  Missing Not at Random}. In \bibinfo{booktitle}{\emph{ACM Conference on
  Knowledge Discovery and Data Mining (KDD)}}. \bibinfo{pages}{713--22}.
\newblock


\bibitem[\protect\citeauthoryear{Steck}{Steck}{2011}]%
        {steck11}
\bibfield{author}{\bibinfo{person}{H. Steck}.} \bibinfo{year}{2011}\natexlab{}.
\newblock \showarticletitle{Item popularity and recommendation accuracy}. In
  \bibinfo{booktitle}{\emph{ACM Conference on Recommender Systems (RecSys)}}.
  \bibinfo{pages}{125--32}.
\newblock


\bibitem[\protect\citeauthoryear{Steck}{Steck}{2019}]%
        {steck19a}
\bibfield{author}{\bibinfo{person}{H. Steck}.} \bibinfo{year}{2019}\natexlab{}.
\newblock \showarticletitle{Embarrassingly Shallow Autoencoders for Sparse
  Data}. In \bibinfo{booktitle}{\emph{International World Wide Web Conference
  (WWW)}}.
\newblock


\bibitem[\protect\citeauthoryear{Tang and Wang}{Tang and Wang}{2018}]%
        {tang17}
\bibfield{author}{\bibinfo{person}{J. Tang} {and} \bibinfo{person}{K. Wang}.}
  \bibinfo{year}{2018}\natexlab{}.
\newblock \showarticletitle{Personalized top-n sequential recommendations via
  convolutional sequence embedding}. In \bibinfo{booktitle}{\emph{ACM
  Conference on Web Search and Data Mining (WSDM)}}.
\newblock


\bibitem[\protect\citeauthoryear{Treister and Turek}{Treister and
  Turek}{2014}]%
        {treister14}
\bibfield{author}{\bibinfo{person}{E. Treister} {and} \bibinfo{person}{J.S.
  Turek}.} \bibinfo{year}{2014}\natexlab{}.
\newblock \showarticletitle{A Block-Coordinate Descent Approach for Large-scale
  Sparse Inverse Covariance Estimation}. In \bibinfo{booktitle}{\emph{Advances
  in Neural Information Processing Systems (NIPS)}}.
\newblock


\bibitem[\protect\citeauthoryear{Verstrepen and Goethals}{Verstrepen and
  Goethals}{2014}]%
        {koen14}
\bibfield{author}{\bibinfo{person}{K. Verstrepen} {and} \bibinfo{person}{B.
  Goethals}.} \bibinfo{year}{2014}\natexlab{}.
\newblock \showarticletitle{Unifying Nearest Neighbors Collaborative
  Filtering}. In \bibinfo{booktitle}{\emph{ACM Conference on Recommender
  Systems (RecSys)}}.
\newblock


\bibitem[\protect\citeauthoryear{Volkovs and Yu}{Volkovs and Yu}{2015}]%
        {volkovs15}
\bibfield{author}{\bibinfo{person}{M.~N. Volkovs} {and} \bibinfo{person}{G.~W.
  Yu}.} \bibinfo{year}{2015}\natexlab{}.
\newblock \showarticletitle{Effective Latent Models for Binary Feedback in
  Recommender Systems}. In \bibinfo{booktitle}{\emph{ACM Conference on Research
  and Development in Information Retrieval (SIGIR)}}.
\newblock


\bibitem[\protect\citeauthoryear{Wiesel and Hero}{Wiesel and Hero}{2012}]%
        {wiesel12}
\bibfield{author}{\bibinfo{person}{A. Wiesel} {and} \bibinfo{person}{A.O.
  Hero}.} \bibinfo{year}{2012}\natexlab{}.
\newblock \showarticletitle{Distributed Covariance Estimation in Gaussian
  Graphical Models}.
\newblock \bibinfo{journal}{\emph{IEEE Transactions on Signal Processing}}
  \bibinfo{volume}{60} (\bibinfo{year}{2012}).
\newblock
Issue 1.


\bibitem[\protect\citeauthoryear{Wu, DuBois, Zheng, and Ester}{Wu
  et~al\mbox{.}}{2016}]%
        {wu16}
\bibfield{author}{\bibinfo{person}{Y. Wu}, \bibinfo{person}{C. DuBois},
  \bibinfo{person}{A.~X. Zheng}, {and} \bibinfo{person}{M. Ester}.}
  \bibinfo{year}{2016}\natexlab{}.
\newblock \showarticletitle{Collaborative Denoising Auto-Encoders for top-N
  Recommender Systems}. In \bibinfo{booktitle}{\emph{ACM Conference on Web
  Search and Data Mining (WSDM)}}.
\newblock


\end{thebibliography}

\end{document}